\documentclass[fleqn,usenatbib]{mnras}

\usepackage{newtxtext,newtxmath}

\usepackage[T1]{fontenc}
\usepackage{ae,aecompl}
\usepackage{enumitem}
\usepackage{diagbox}

\usepackage{url}
\usepackage{hyperref}

\newlist{inparaenum}{enumerate}{1}
\setlist[inparaenum]{nosep}
\setlist[inparaenum,1]{label=1\alph*)}

\newlist{num}{enumerate}{1}
\setlist[num]{nosep}
\setlist[num,1]{label=\arabic{numi})}

\newcommand{\ha}{{H$\alpha$}}
\newcommand{\gaia}{\textit{Gaia}}

\newcommand{\iraf}{IRAF}
\newcommand{\gcmd}{\textit{Gaia} CMD}
\newcommand{\ucmd}{UVEX CMD}
\newcommand{\wccd}{WISE CCD}
\newcommand{\iccd}{IPHAS CCD}

\usepackage{graphicx}	
\usepackage{amsmath}	

\usepackage{mathtools}

\usepackage{array}
\usepackage{multirow}
\usepackage{lscape}
\usepackage{hyperref}
\usepackage{caption}
\usepackage{subcaption}
\usepackage{longtable}
\usepackage{supertabular,booktabs}

\title[Spectroscopic validation of the \gaia/IPHAS catalogue.]{Spectroscopic follow-up of a sub-set of the \gaia/IPHAS catalogue of \ha-excess sources.}

\author[Fratta et al.]{M. Fratta,$^{1}$\thanks{E-mail: matteo.fratta@durham.ac.uk}
S. Scaringi,$^{1}$
M. Mongui\'o,$^{2,3,4}$
A. F. Pala,$^{5}$
J. E. Drew,$^{6}$
C. Knigge.$^{7}$
K. A. I\l{}kiewicz,$^{1}$
P. Gandhi,$^{7}$
\\
$^{1}$Centre for Extragalactic Astronomy, Department of Physics, University of Durham, South Road, Durham, DH1 3LE, UK\\
$^{2}$Institut de Ci\`encies del Cosmos (ICCUB), Universitat de Barcelona (UB), Mart\'i i Franqu\`es 1, E-08028 Barcelona, Spain\\
$^{3}$Departament de Física Qu\`antica i Astrof\'isica (FQA), Universitat de Barcelona (UB), Mart\'i i Franqu\`es 1, E-08028 Barcelona, Spain\\
$^{4}$Institut d'Estudis Espacials de Catalunya (IEEC), c. Gran Capit\`a, 2-4, 08034 Barcelona, Spain\\
$^{5}$European Space Agency, European Space Astronomy Centre, Camino Bajo del Castillo s/n, 28692 Villanueva de la Cañada, Madrid, Spain\\
$^{6}$Dept of Physics and Astronomy, Faculty of Maths and Physical Sciences, University College London, Gower Street, London, WC1E 6BT, UK\\
$^{7}$School of Physics and Astronomy, University of Southampton, University Road, Southampton, SO17 1BJ, UK\\
}

\date{Accepted XXX. Received YYY; in original form ZZZ}

\pubyear{2022}

\usepackage{etoolbox}

\begin{document}
\label{firstpage}
\pagerange{\pageref{firstpage}--\pageref{lastpage}}
\maketitle

\begin{abstract}
\leftskip=133pt
State-of-the-art techniques to identify \ha\ emission line sources in narrow-band photometric surveys consist of searching for \ha\ excess with reference to nearby objects in the sky (position-based selection). However, while this approach usually yields very few spurious detections, it may fail to select intrinsically faint and/or rare \ha-excess sources. In order to obtain a more complete representation of the heterogeneous emission line populations, we recently developed a technique to find outliers relative to nearby objects in the colour-magnitude diagram (CMD-based selection). By combining position-based and CMD-based selections, we built an updated catalogue of \ha-excess candidates in the northern Galactic Plane. Here we present spectroscopic follow-up observations and classification of 114 objects from this catalogue, that enable us to test our novel selection method. Out of the 70 spectroscopically confirmed \ha\ emitters in our sample, 15 were identified only by the CMD-based selection, and would have been thus missed by the classic position-based technique. In addition, we explore the distribution of our spectroscopically confirmed emitters in the \gcmd. This information can support the classification of emission line sources in large surveys, such as the upcoming WEAVE and 4MOST, especially if augmented with the introduction of other colours.
\end{abstract}

\begin{keywords}
\leftskip=133pt
Stars: emission-line - techniques: spectroscopic - techniques: photometric
\end{keywords}



\section{Introduction}
\label{sec: introduction}
Unresolved \ha\ emission can be associated with very heterogeneous populations of both single and binary stars, spanning varied evolutionary stages. This diversity makes the study of emission line sources particularly important to understand the composition and evolution of our Galaxy. While some of these objects are frequently observed, others might be intrinsically very faint and/or belong to the rarest types of stellar populations. Isolated \ha-emitting sources include young stellar objects (YSOs), coronally active M-Dwarfs, luminous blue variables (LBVs), classical Be stars, and Wolf-Rayet stars. Accretion driven \ha\ emission can also be observed from interacting binaries, such as cataclysmic variables (CVs), symbiotic stars (SySts), and accreting neutron stars or black holes. This types of sources are not only rare, but also difficult to both identify and precisely classify\footnote{For instance, the list of CVs produced by \cite{ritter} includes 1,429 objects.}.
Even more challenging is the classification of objects that may not display clear \ha-emission line properties, but rather display \ha-excess relative to their parent population. Even though these source types do not present evidence of \ha\ emission, their intensities in said band might still significantly exceed the ones of the objects they are compared to. An example is provided by the elusive population of low mass transfer rate CVs that have evolved beyond the so-called period minimum (\citealt{per_minimum1}; \citealt{per_minimum2}; \citealt{per_minimum3}).
Among the several ways to identify and classify stellar objects, the most telling is a moderate-resolution spectroscopy. However, spectra collection tends to be generally expensive, in particular when attempting to classify large ensembles of targets. Similar issues also affect classification studies based on light-curve analysis.

In order to alleviate the burden, several pre-selection methods have been developed and applied over the years. A popular approach to pre-select targets is based on optical photometry; it deploys specific colour and magnitude cuts to select targets for spectroscopy. Specifically related to \ha-emission line sources, \cite{witham} leveraged photometric measurements from the INT Photometric \ha\ Survey of the Northern Galactic Plane (IPHAS; \citealt{iphas}) to produce a list of $\sim 5,000$ \ha-excess candidates. These objects were selected by isolating \ha-bright targets in the r-\ha\ vs. r-i colour-colour diagram (CCD), on a field-by-field basis. More recently, \cite{monguio} produced the IGAPS catalogue, thus expanding and refining the work of \cite{witham}.
By testing $\sim 53,000,000$ targets for \ha\ excess, they identified $\sim 21,000$ excess-line candidates relative to spatially nearby sources in the sky, from IPHAS and UVEX (UV-Excess survey of the Northern Galactic Plane; \citealt{uvex}) surveys. These ``position-based'' selections are generally rather conservative, and the amount of bogus detections is usually low. While \cite{monguio} did not pursue a spectroscopic confirmation of their selection, \cite{witham} observed the spectra of $\sim 300$ of their \ha-excess candidates. Out of them, 97\% were thus confirmed as emission-line sources. However, position-based approaches, due to their conservative nature, may fail to identify some outliers.

In order to include these objects, \cite{fratta} (henceforth FR21) developed a novel technique that prioritises a better representation of the potential \ha-emitting populations. This improvement can sometimes be achieved at the expense of a higher false-positive rate. Taking advantage of the \gaia\ DR2 parallaxes, their method consisted in augmenting the classic position-based selection by identifying \ha-excess candidates relative to nearby sources in the M\textsubscript{G} vs. G\textsubscript{BP} - G\textsubscript{RP} colour-absolute\footnote{To simplify the notation, in the current paper the adjective ``absolute'' is implicit.} magnitude diagram (CMD-based selection). However, it is relevant to point out that the absolute magnitudes in the G band, as well as the colour indices, were calculated with the assumption of zero extinction, and therefore they should be treated as upper limits. This contributes to increase the effects of population mixing in this parameter space. 
The authors assigned two \textit{significance values} (or simply ``\textit{significances}''\footnote{In FR21, the authors refer to the \textit{significance} just as $\sigma$. As $\sigma$ is commonly associated to the standard deviation of a distribution, a different notation is preferred in the current paper.}, n\textsubscript{$\sigma$}) of \ha-excess to each object in their meta-catalogue, one for each partitioning criterion (position-based or CMD-based): these defined the distance from the main stellar locus in IPHAS r-\ha\ vs. r-i parameter space, expressed in units of standard deviation of the r-\ha\ distribution.
The objects with at least one significance higher than three were identified as ($3\sigma$) \ha-excess candidates. Out of $\sim$7.5\,million targets, this selection criterion yielded $\sim$28,000 \ha-excess candidates.
To test the statistics of their selection, FR21 visually inspected $\sim2,000$ randomly selected spectra from LAMOST DR5 database \citep{lamost}, relative to a subset of their $3\sigma$ outliers. Among these spectra, 49\% showed \ha\ emission; a more conservative 5$\sigma$ threshold yielded $\sim$7,000 outliers, and the \textit{purity} of this cut was 82\%. However, we point out that LAMOST observation strategy privileges relatively blue, bright objects in the Galaxy \citep{selection_lamost}. Therefore, an unconstrained cross-match with this data does not necessarily result in a homogeneous representation of the objects in the \gcmd.

This study is part of a pilot program to validate and test the CMD-based technique described in FR21, while simultaneously identifying and classifying new \ha-excess sources. To this end, a set of spectroscopic follow-up observations of the brightest $5\sigma$ \ha-excess candidates from FR21 is presented. 
The targets examined here are chosen in an attempt to homogeneously sample the \gcmd.
In addition, we discuss in Appendix \ref{sec: additional} how the distribution of our spectroscopically confirmed \ha-emitters across the \gcmd, supplemented by UVEX and WISE parameter spaces, can be used to efficiently separate the different populations. 
This is essential to ease the task of target selection for large surveys, such as the forthcoming WEAVE multi-object fibers facility (\citealt{weave1}; \citealt{weave2}), or 4MOST \citep{4most}. With its $\sim$1,000 fibers, its large Integral Field Unit (IFU), and 20 mini IFUs, WEAVE will provide spectroscopic follow-up observations for \gaia\ and for the Low Frequency Array (LOFAR; \citealt{lofar}) surveys, in the 370-1000\,nm wavelength range, with intermediate (R$\sim5,000$) and high resolution (R$\sim20,000$). The 4-metre Multi-Object Spectroscopic Telescope (4MOST) survey will be able to simultaneously obtain spectra for $\sim2,400$ objects, with a resolution range comparable to WEAVE.

\smallskip

The paper is organised as follows: Section \ref{sec: input} introduces the initial selection cuts employed to define our target list. Gemini spectroscopic observations and data reduction are also discussed. In Section \ref{sec: classification}, our source classification is presented. In Section \ref{sec: validation} our spectroscopic classification is used to test the CMD-selection by FR21. In Section \ref{sec: additional}, additional photometric parameter spaces for stellar population discrimination purposes are explored. In Section \ref{sec: conclusion}, our conclusions are drawn.

\section{Observations and Data Reduction}
\label{sec: input}

The $\sim28,000$ $3\sigma$ \ha-excess candidates identified by FR21 are located in the Northern Galactic Plane, within the $|b|\leq 5$\textdegree\ and 29\textdegree $\leq l \leq$ 215\textdegree\ coordinate ranges, and inside a 1.5\,kpc radius relative to the Sun\footnote{However, we point out that this threshold does not constitute a hard cut, and the distance of some objects from the Sun exceeds 1.5\,kpc.}.
The distances were calculated with the parallax-inversion method; the ``$\pi/{\sigma_{\pi}} > 5$'' constraint (with $\pi$ representing the parallax, and $\sigma_{\pi}$ the corresponding uncertainty) for parallax measurements minimised the difference between this estimator and the distances obtained with a probabilistic approach \citep{dist}. The availability of \gaia\ DR2 parallaxes enabled FR21 to locate the objects in the CMD, and thus complement the classic position-based \ha-excess candidates selection with the novel CMD-based technique. The results described in FR21 can be updated with the use of more precise \gaia\ DR3 parallaxes. However, the average difference between parallaxes from \gaia\ DR2 and EDR3 is 0.06\,mas, for all the targets inspected by FR21.

\begin{figure}
    \centering
	\includegraphics[width=\columnwidth, trim= 0mm 0mm 0mm 0mm]{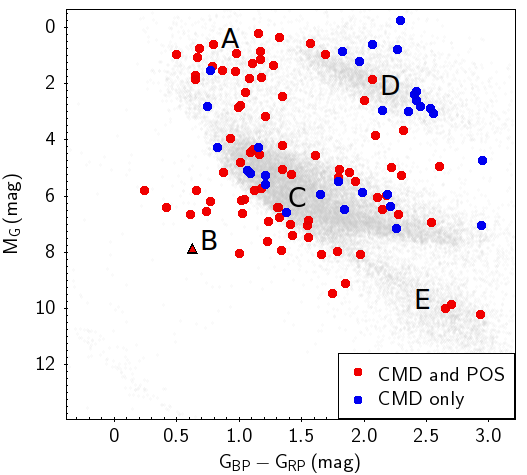}
	\includegraphics[width=\columnwidth, trim= 2mm 0mm 0mm 0mm]{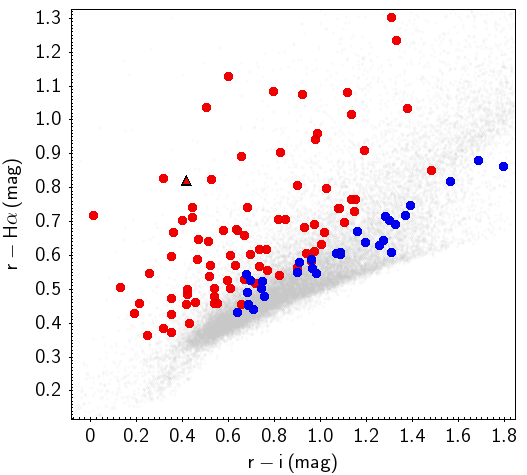}
    \caption{Locations in the \gcmd\ and in the \iccd\ of the 114 sources in our dataset (top and bottom panels, respectively). While red colour refers to the \ha-excess candidates selected in FR21 from both partition types, the targets depicted in blue were identified from CMD-partitions only. No objects in our sample were identified in positional-partitions only. The grey dots plotted in the background as a reference represent all the $3\sigma$ outliers selected by FR21. The black-red triangle refers to the object \gaia\ DR2 259018688664493312, mentioned in the text.}
    \label{fig:sources}
\end{figure}

The \ha-excess candidates from FR21 are represented by the grey dots in Fig.~\ref{fig:sources}. From their distribution in the \gcmd\ (top panel), they can be tentatively visually divided into five groups. The regions of this parameter space occupied by each group are labelled with capital letters. Region A is usually associated to bright and relatively blue Be/Ae stars; CVs mostly populate region B, i.e. the area between the Zero Age Main Sequence (ZAMS) and the white dwarf (WD) tracks; the targets in region C are mainly pre-Main Sequence objects; D area is a predominantly populated by reddened red giants (RGs), but also reddened Be stars and bright YSOs can be found; E region is commonly associated to faint M-Dwarfs. However, a classification based on this criterion is indicative at best, as it is strongly affected by population mixing effects, and by the zero-extinction assumption.

\subsection{Sample Selection}
\label{sec: sample}

Our spectroscopic data were obtained at the Gemini (North and South) telescopes, via a poor-weather queue. In order to reach a compromise between a minimum required signal-to-noise ratio of 10 (even for the faintest objects), and the necessary observing time, only objects with $r<17.5$\,mag were included in our follow-up. Only targets with a significance of being \ha-excess sources higher than or equal to 5 were considered.
In an attempt to obtain a homogeneous sample across the tentative clusters A-E, our strategy has been to request follow-up observations of up to 2,230 targets, sampled as evenly as possible across the the \gcmd. Out of these 2,230 spectra, 114 were obtained (67 from Gemini North, 47 from Gemini South).
These are displayed as the red and blue dots in the \gcmd\ and in the \iccd\ (top and bottom panels, respectively). The red candidates were identified by both position-based and CMD-based outliers selection, while the blue ones were detected only by the CMD-based selection. None of these outliers resulted only from the position-based selection. As it can be seen in the top panel, the minimum-brightness constraint excludes from our spectral follow-up all the \ha-excess candidates that crowd the lowest stripe of the \gcmd. Among them we can list, for instance, all the objects that lie on (or around) the WD track.
The bottom panel in the same figure brings out the value of the CMD-based selection technique. In fact, a non-negligible amount of outliers from FR21 (mainly blue dots) blends with the main stellar locus, in the IPHAS parameter space; without the prior population discrimination, these objects would have been hidden.

\subsection{Gemini spectra}
The spectra analysed in this work were acquired with the Gemini Multi-Object Spectrograph (GMOS; \citealt{gmos2}; \citealt{gmos1}), with individual exposure times of 450\,s. Gemini observatory consists of two 8.1\,m telescopes, based in Mauna Kea (Hawaii; Gemini North) and in Cerro Pachon (Chile; Gemini South), respectively. The observations were carried out between August 2019 and the end of January 2020. The use of the R400 grating yielded a spectral resolution at the blaze wavelength (764\,nm) of R$\sim1,900$. The obtained wavelength resolution is 1\,\AA.

\begin{figure}
	\includegraphics[width=\columnwidth, trim= 0mm 0mm 0mm 0mm]{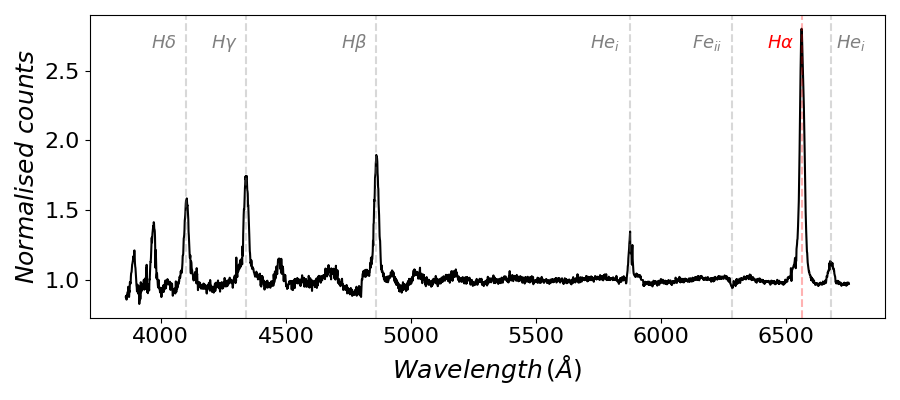}
    \caption{Example of reduced Gemini spectrum, belonging to the object \gaia\ DR2 259018688664493312. The most prominent emission/absorption lines are highlighted by dashed vertical lines.}
    \label{fig:example_spectrum}
\end{figure}

The setup of the detector caused systematic gaps in the data. To eliminate this effect, two spectra with different central wavelengths (more specifically, 525 and 535\,nm, respectively) were consecutively collected for each target\footnote{For a total observation time of 900\,s per target.}. These two spectra were then averaged, in order to associate one single spectrum to each source in our list.
Gemini raw data were reduced with the Image Reduction and Analysis Facility Software System (\iraf; \citealt{iraf}) and the SAOImage DS9 tool (DS9; \citealt{ds9}). The process involved a cosmic rays subtraction, as well as bias and flat-field corrections. CuAr arcs were used during wavelength calibration step. Flux calibration could not be performed, because the poor-weather program did not include the observation of any flux standard objects. A spline function was instead fitted to the non-calibrated counts. During this procedure, the spectral lines were masked-out. The flux was then divided by the spline.
Fig.~\ref{fig:example_spectrum} displays an example of a reduced spectrum. It belongs to the object \gaia\ DR2 259018688664493312, the position of which is highlighted with a red triangle in Fig.~\ref{fig:sources}.

\section{Sources classification}
\label{sec: classification}

While testing the selection technique presented in FR21, our spectral analysis also aims to break the degeneracy that emerges from tentatively characterising the sources on the basis of their location in the \gcmd. Our classification is made more solid with the support of previous designations found in SIMBAD \citep{simbad} database (when available). Out of 114 targets, 32 are already labelled in SIMBAD. Our classifications confirm the ones found in said database for almost half of the matches (15/32). In the remaining 17 cases, SIMBAD does not provide unambiguous classifications: while 12 of these targets are classified as ``emission line stars", 4 are labelled as ``variable stars", and one simply as ``star".

The most common spectral features that distinguish the stellar populations mentioned in Section \ref{sec: input} are the following. Given the distribution of our targets in the \gcmd, Main Sequence (MS) stars other than classical Be stars and active M-dwarfs are also listed here:
\begin{itemize}
    \item \textbf{MS stars (except for classical Be stars and active M-dwarfs)}: their spectra are characterised by absorption in the main Balmer and He lines (the latter ones are shown mainly in the hottest objects of this category). Among them, $\gamma$-Doradus are typically A-F spectral types stars, characterised by line-profile and radial velocities variability (predominantly produced by g-mode oscillations; \citealt{gamma});
    \item \textbf{coronally-active M-dwarfs}: molecular (TiO and CaH) absorption is one of the most prominent spectral features of (mainly early type) M-dwarfs (\citealt{md_features1}; \citealt{md_features2}) due to the low atmosphere temperature. This population usually shows relatively narrow \ha\ (and sometimes Ca) emission lines due to coronal activity;
    \item \textbf{Be and Ae stars}: the spectra of both these object types are characterised by a blue continuum, with narrow \ha\ emission. Broad absorption lines might be found in correspondence to the other Balmer lines. Furthermore, Fe\textsubscript{II} and O\textsubscript{I} emission lines might be observed in the spectra of Be stars, while Ae stars often show emission in some forbidden lines. One of the peculiarities of these classes of objects is the high variability of the strength of the emission lines \citep{be_features};
    \item \textbf{RGs}: as red giants, these objects are characterised by a red continuum emission. Narrow absorption is commonly observed in the \ha, He and Fe\textsubscript{II} lines;
    \item \textbf{YSOs}: due to the diverse nature of the Pre-Main-Sequence objects, their spectra can show very different features. Nonetheless, they usually present an IR excess, as well as the \ha\ line in emission. Additionally, TiO and VO molecular absorption bands can be often spotted from T-Tauri stars and from Orion variables \citep{yso_features}. Sources in this latter group might also present Paschen emission lines in their spectra \citep{hillenbrand};
    \item \textbf{CVs}: generally characterised by a blue continuum; broad Balmer - and often He - emission lines originate from the hot accretion disk (\citealt{cv_features1}; \citealt{cv_features2}).
\end{itemize}

\begin{figure}
	\includegraphics[width=\columnwidth, trim= 0mm 0mm 0mm 0mm]{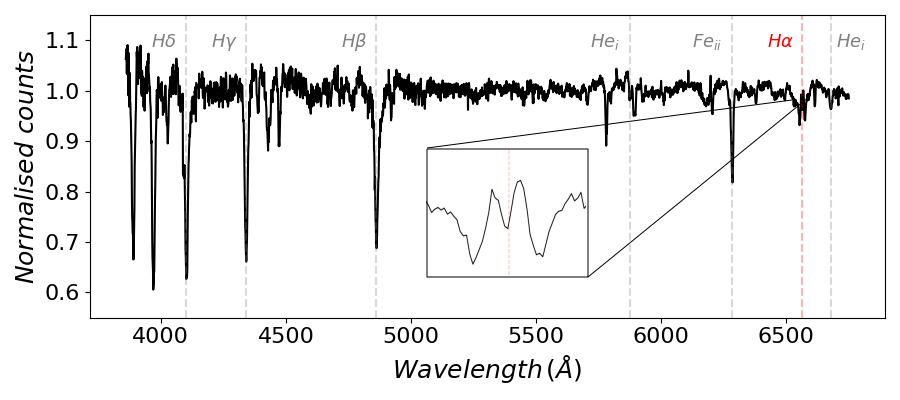}
	\caption{Spectrum of the object \gaia\ DR2 3455444577118544768. The shape in correspondence of the \ha\ wavelength does not allow the measurement of its FWHM.}
    \label{fig:weird_spectra}
\end{figure}

\begin{figure}
	\includegraphics[width=\columnwidth, trim= 0mm 0mm 0mm 0mm]{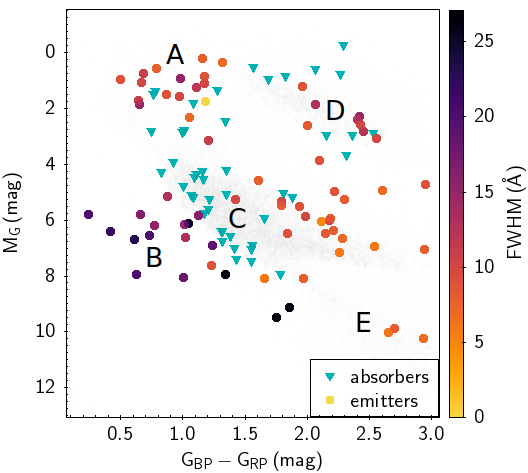}
    \caption{Positions of our targets in \gaia\ CMD, colour-coded according to the FWHM of the \ha\ line. The objects that show \ha\ absorption are depicted by the turquoise triangles. The grey dots represent all the $3\sigma$ outliers selected by FR21.}
    \label{fig:FWHM_gaia}
\end{figure}

Target classification can be further supported by the analysis of the full width at half maximum (FWHM) of the \ha\ line. Henceforth, the acronym ``FWHM'' will refer to the full width at half maximum of the \ha\ line, unless otherwise specified. The width of a given line can provide an indication for the presence of an accretion disk (in compact accretors): due to Doppler broadening, disk-fed accretors are characterised by relatively broad spectral lines. 
Out of the inspected spectra, 69/114 belong to confirmed \ha\ emitters, and the FWHM is measured for 68 of them. The peculiar shape of the \ha\ line of one of our sources does not allow the measurement of its FWHM, as testified by Fig.~\ref{fig:weird_spectra}. The distribution of our targets in the \gcmd, colour-coded with respect to their FWHM, is displayed in Fig.~\ref{fig:FWHM_gaia}.
The \ha\ emitters (the coloured dots), are evenly spread on the CMD. The objects with the largest FWHM can be found between the ZAMS and the WD tracks. Said region of the \gcmd\ is commonly associated to accreting WDs; both these source types are characterised by the presence of a disk surrounding the accreting object (\citealt{cv_disk1}; \citealt{cv_disk2}, \citealt{yso_disk1}; \citealt{yso_disk2}). The remaining 45 objects, characterised by absorption in the \ha\ line, are depicted by the turquoise triangles. These objects are discussed later in the text.

\smallskip

\begin{figure*}
	\includegraphics[width=\textwidth, trim = 0mm 0mm 0mm 0mm]{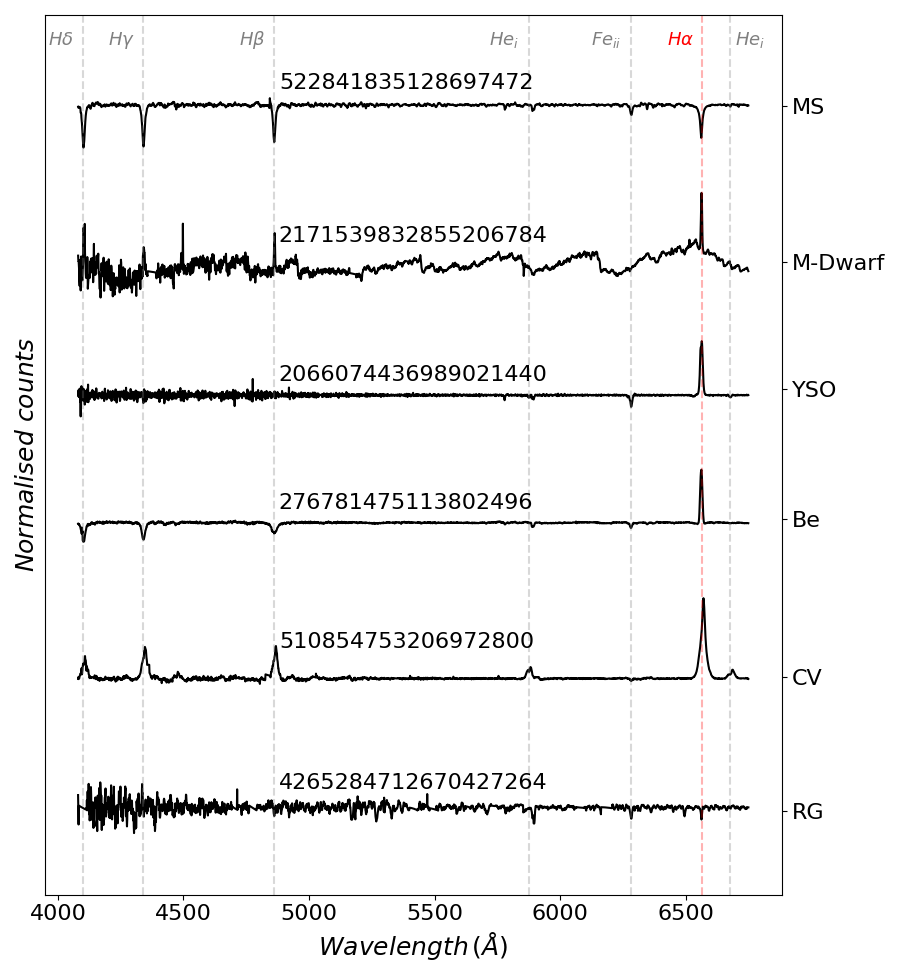}
    \caption{The spectra shown here (flux-normalised for visualisation purposes) exemplify the six categories to which the objects in our sample are assigned. The labels on the right vertical axis associate each spectrum to its corresponding population. The respective targets are identified by their (\gaia\ DR2) Source ID. Some Balmer, He and Fe lines are highlighted with dotted lines.}
    \label{fig:example_spectra}
\end{figure*}

\begin{figure*}
	\includegraphics[width=5\textwidth/6, trim= 0mm 0mm 0mm 0mm]{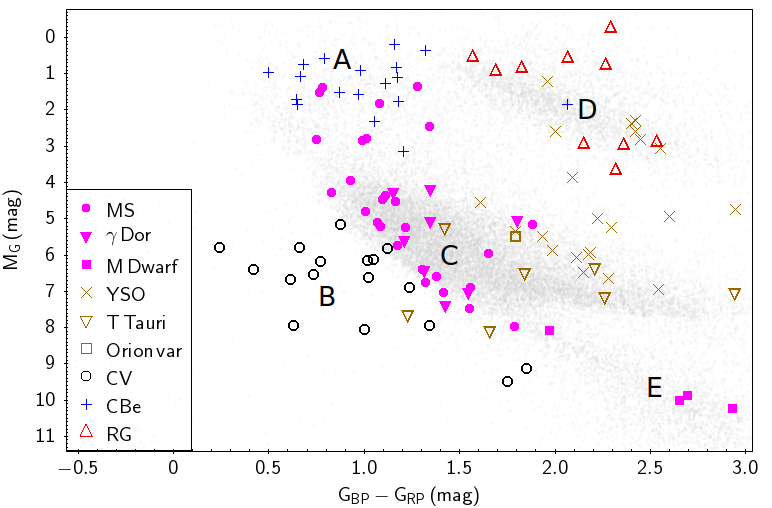}
	\caption{Location of our sources in the \gcmd, colour-coded with respect to their spectral classification. The grey dots represent the \ha-excess candidates selected in FR21.}
    \label{fig:classification_gaia}
\end{figure*}

As an outcome of our spectral classification, the examined dataset is found to be a compound of: 35 MS stars (except for Be stars and M-dwarfs); 30 YSOs (while \citealt{Marton} recently produced a catalogue of more than 1.1 millions of YSO candidates, selected from \gaia\ DR2 and WISE surveys with machine learning techniques); 18 Be stars (whereas the all-sky Be Star Spectra (BeSS) database described in \citealt{neiner} counts more than 2,000 Be stars); 17 CVs (to be added to the 1,429 CVs listed by \citealt{ritter}); 10 RGs; 4 active M-dwarfs (\citealt{lepin} found $\sim$9,000 bright active M-dwarfs from the SUPERBLINK survey. However, the authors estimate that their selection is only $\sim$75\% complete).
Fig.~\ref{fig:example_spectra} shows six exemplary spectra, one for each identified population. Our results are summarised in Table~\ref{tab: results}, in the appendix. 
The taxonomy of our sample is shown in Fig.~\ref{fig:classification_gaia}, where the various populations are represented in the \gcmd\ with different colours and shapes.
As can be noticed, targets labelled likewise generally tend to cluster in well defined regions of the \gcmd\ - the only noticable exception to this statement is provided by YSOs which, for their manifold nature, are more widely spread in this parameter space.
However, the overlaps of these regions clearly demonstrate that the location in the \gcmd\ does not provide enough information to unambiguously tie a source to a stellar population. For instance: a particularly dominant companion would make it an extremely difficult task to discern a CV from a MS star; moreover, the bright and red tail of the YSO distribution seems to blend with the RGs; furthermore, distinguishing line emitting Be stars from common, bright, MS stars can be very tricky.

\section{Testing the CMD-based identification}
\label{sec: validation}

\begin{figure}
    \centering
	\includegraphics[width=\columnwidth, trim= 0mm 0mm 0mm 0mm]{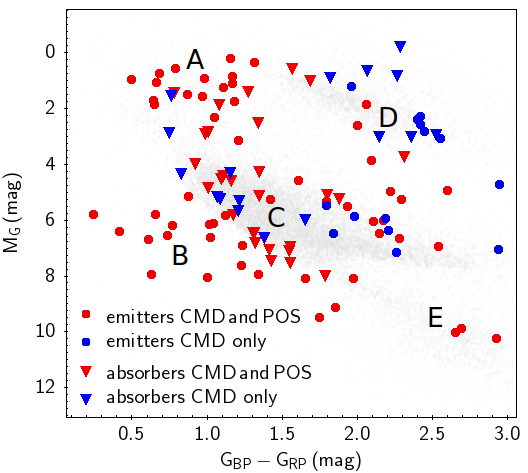}
	\includegraphics[width=\columnwidth, trim= 1mm 0mm 2mm 0mm]{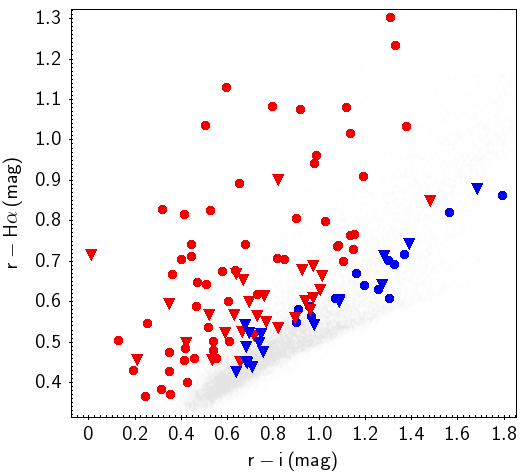}
    \caption{Locations in the \gcmd\ and in the \iccd\ of the 114 sources in our dataset (top and bottom panels, respectively). The colour-code is the same of Fig.~\ref{fig:sources}. While the dots display the \ha\ line in emission, the triangles show it in absorption.}
    \label{fig:sources_em_abs}
\end{figure}

The distribution of our sample in the \gcmd\ (top panel) and in the \iccd\ (bottom panel) is shown in Fig.~\ref{fig:sources_em_abs}. The colour-code is the same applied in Fig.~\ref{fig:sources}. Here the dots represent the spectroscopically confirmed \ha\ emitters, while the triangles display the absorbers. Out of the 69 emitters, 55 were selected by both the position-based and CMD-based techniques. The remaining 14 emitters were obtained only through the CMD-based method (no outliers in our sample were identified by the position-based selection only), demonstrating the improvements introduced by this selection in terms of completeness. This is further shown by the fact that the CMD-significance is higher than the position-significance for 45/55 of the common emitters.
All the 14 CMD-emitters are spectroscopically classified as YSOs. This confirms that the CMD-based method is more efficient than the position-based one in identifying emission-line sources within this population (as also stated in Section 5 of FR21).

Out of the 45 absorbers, 28 (62\%) were selected by both position-based and CMD-based selections. The remaining 17 absorbers were identified only by the CMD-based method.
The 45 absorbers include 35 MS stars (other than classical Be stars and M-dwarfs) and all the 10 RGs. The variability timescale typical of these stellar populations is significantly longer than the $\sim10$\,years that separate IPHAS and LAMOST observations. For this reason, the photometric identification of these absorbers as \ha-excess candidates cannot be justified solely on the basis of temporal variability. However, it is relevant to stress that the algorithm presented in FR21 does not aim at identifying \ha-\textit{emitters}, but rather objects that present an \textit{excess} in the \ha\ flux, with reference to nearby targets in the \gcmd. Fig. 17 in FR21 shows two examples to visually explain this phenomenon.
Out of the 10 RGs in our sample, 7 were selected only by the CMD-based method. To test the reliability of this selection criterion in region D of the CMD, a group of outliers that lie in that area is visually isolated. Considering that the vast majority of them was identified only by the CMD-based method, the purity yielded by this selection criterion in region D of the \gcmd\ seems to be significantly lower than the one from the position-based selection. However, half of the YSOs identified only by the CMD-based method also lie in that area. The position-based selection alone would have missed these outliers.
A similar test is performed for the MS stars population: the majority of the visually selected outliers along the ZAMS were identified only by the CMD-based technique. However, 25/35 of the MS stars in our sample were identified by both selections. Therefore, the position-based one does not seem to provide advantages in terms of purity, along the ZAMS. On the other hand, one YSO that lies region was selected only by the CMD-based technique.

Table~\ref{tab: stats} summarises how many members of each population in our sample were identified as \ha-excess candidates by the CMD-based selection and how many by the position-based one. The number of spectroscopically identified \ha-absorbers in each population is provided within the parentheses.

\begin{table}\centering
\begin{tabular}{|c|c|c|}
 \hline
  
 \shortstack{{} \\ Population \\ {}} & \shortstack{CMD-based \\ selection (abs)} & \shortstack{position-based \\ selection (abs)} \\
  \hline
  YSO & 30 (0) & 16 (0)\\
  \hline
  Classical Be & 18 (0) & 18 (0) \\
  \hline
  CV & 17 (0) & 17 (0) \\
  \hline
  RG & 10 (10) & 3 (3) \\
  \hline
  Active M-dwarf & 4 (0) & 4 (0) \\
  \hline
  Other MS star & 35 (35) & 25 (25) \\
  \hline
  Total & 114 (45) & 83 (28) \\
  \hline
  
 \end{tabular}
 \caption{Number of members from each population in our sample detected by the CMD/position-based selections, respectively. The amount of absorbers from each population is included inside the parentheses.}
 \label{tab: stats}
\end{table}


\section{Conclusions}
\label{sec: conclusion}

\cite{fratta} developed a method to identify \ha-excess candidates with reference to nearby sources in the \gcmd. In this study, part of a pilot-program to validate this CMD-based selection technique is presented. More specifically, its performances in terms of \textit{completeness} are compared to those of classical position-based methods. 
Our analysis involves the spectral observation and classification of 114 bright ($r\leq17.5$\,mag) \ha-excess candidates. These are drawn from a group of 2,230 $5\sigma$ outliers that homogeneously sample the \gcmd. Their spectra were obtained by the two Gemini telescopes. 
Table \ref{tab: stats} shows how many \ha-excess candidates were identified by each selection criterion, on a population-by-population basis. The amount of \ha-absorbers in each group of objects is specified within parentheses.
Since the spectra that present the \ha\ line in absorption belong to either RGs or MS stars (other than classical Be stars or active M-dwarfs), \ha\ emission variability seems to be discarded as a possible explanation for their detection as outliers.
Supplementary analyses to justify the selection of these objects as \ha-excess sources are required.

We further show that the positions of our spectroscopically classified objects in the \gcmd, \ucmd\ and \wccd\ can be used to better discern the populations of \ha\ emission-line sources. A pre-selection based on this criterion can be used to improve the identification and classification of emission-lines sources in large databases. This is a particularly relevant task, specially with the upcoming arrival of large spectroscopic surveys, such as WEAVE and 4MOST.

\section{Data availability}
The 114 analysed Gemini spectra can be found in VizieR as ``Spectroscopic follow-up of \ha-excess sources in the Northern Galactic Plane''.



\section*{Acknowledgements}
Based on observations obtained at the international Gemini Observatory, a program of NSF’s NOIRLab, which is managed by the Association of Universities for Research in Astronomy (AURA) under a cooperative agreement with the National Science Foundation, on behalf of the Gemini Observatory partnership: the National Science Foundation (United States), National Research Council (Canada), Agencia Nacional de Investigaci\'{o}n y Desarrollo (Chile), Ministerio de Ciencia, Tecnolog\'{i}a e Innovaci\'{o}n (Argentina), Minist\'{e}rio da Ci\^{e}ncia, Tecnologia, Inova\c{c}\~{o}es e Comunica\c{c}\~{o}es (Brazil), and Korea Astronomy and Space Science Institute (Republic of Korea).
Acquired through the Gemini Observatory Archive at NSF’s NOIRLab* and processed using the Gemini IRAF package.
Program IDs: GN-2019B-Q-401 and GS-2019B-Q-403.

\smallskip

MM work was funded by the Spanish MICIN/AEI/10.13039/501100011033 and by ``ERDF A way of making Europe'' by the ``European Union'' through grant RTI2018-095076-B-C21, and the Institute of Cosmos Sciences University of Barcelona (ICCUB, Unidad de Excelencia ’Mar\'{\i}a de Maeztu’) through grant CEX2019-000918-M.

\smallskip

This research has made use of the SIMBAD database,
operated at CDS, Strasbourg, France \citep{simbad}.

\bibliographystyle{mnras}
\bibliography{refs.bib} 




\appendix



\section{Photometry-based pre-selection}
\label{sec: additional}
The availability of large spectroscopic surveys makes the exploration of different pre-selection methods a crucial task. The last paragraph of Section \ref{sec: classification} shows how the distribution of our sample in the \gcmd\ can be used as indication to classify \ha-excess sources. The current study shows how this information can be augmented with additional diagnostics, to ease a spectroscopic classification. The physical and chemical properties of diverse source types are can affect wider different pass-bands out of the \gaia\ ones. Additional information from bluer/redder pass-bands can thus better inform on the exact class.

The optical photometry from \gaia\ is complemented here with UV and IR intensities. These are leveraged from UVEX and WISE surveys, respectively. The UV-Excess survey of the Northern Galactic Plane (UVEX) started in 2006, and it is based on the island of La Palma (Spain). With the same set-up as IPHAS, the Wide Field Camera (WFC) mounted on the Isaac Newton Telescope (INT) provides U\footnote{We point out that U band measurements in the IGAPS catalogue are not uniformly calibrated.}, g, r and He\textsubscript{I5875} magnitudes for targets in $|b|\leq 5$\textdegree\ (\citealt{uvex}; \citealt{monguio}), up to 22\,mag. Its main task consists in the identification of new Galactic stellar remnants.
On the other hand, the Wide-field Infrared Survey Explorer (WISE) is an all-sky survey funded by NASA, that was launched in December 2009. In six months, it scrutinises the IR properties of the entire sky in four bands (W1, W2, W3, and W4), centred at 3.4, 4.6, 12, and 22\,$\mu$m \citep{wise}. With its four million pixels, this survey achieves a remarkable sensitivity: more than 1,000 times better than IRAS (InfraRed Astronomical Satellite; \citealt{iras1}) in the 12\,$\mu$m band, and 50,000 times better than DIRBE (Diffuse InfraRed Background Experiment; \citealt{dirbe}) in the two mid-infrared bands. Thanks to its features, WISE is particularly efficient in detecting (even old, faint) Brown Dwarfs (BDs), Ultra-Luminous InfraRed Galaxies (ULIRGs), Active Galactic Nuclei (AGNs) and Quasi Stellar Objects (QSOs).

\smallskip

Out of the 114 \ha-excess candidates in our dataset, 78 have an available measurement in UVEX U, g and r bands, whilst 73 have associated W1, W2, and W3 WISE measurements\footnote{We point out that the large sizes of WISE pixels might produce some spurious cross-matches, specially in crowded fields such as the one being studied here.}. A cross-match between these two groups yields 57 matches, i.e. half of our dataset. The non-detection of a source carries valuable information for classification purposes, per se. In fact, objects that appear to be too faint (or too bright), in relation to the limiting magnitudes of the survey, are not detected.
Among the 36 objects in our sample that are not included in UVEX survey, 27 are MS stars\footnote{However, it is worth noticing that all the four M-Dwarfs in our sample are included in UVEX catalogue.} (one of them being a Be star), and eight are RGs. These two populations are usually faint in UVEX bands, as confirmed by the fact that the MS stars and RGs included in UVEX are among the faintest objects in our sample, in the U band. The faintest target in our sample sets a lower limit of U\textsubscript{MS/RG}>21.3\,mag on the apparent magnitude of the remaining MS stars and RGs in our sample.
Similarly, 23 MS stars (including three Be stars and one M-dwarf) and 13 CVs have no detections in WISE bands. The sources in these categories with also a WISE counterpart are among the faintest in all WISE bands of interest. The W1\textsubscript{MS/CV}>14.8\,mag, W2\textsubscript{MS/CV}>14.8\,mag, and W3\textsubscript{MS/CV}>12.7\,mag constraints for the remaining MS stars and CVs in our sample are set by the faintest objects with WISE counterparts.

\begin{figure}
    \centering
	\includegraphics[width=\columnwidth, trim= 0mm 0mm 0mm 0mm]{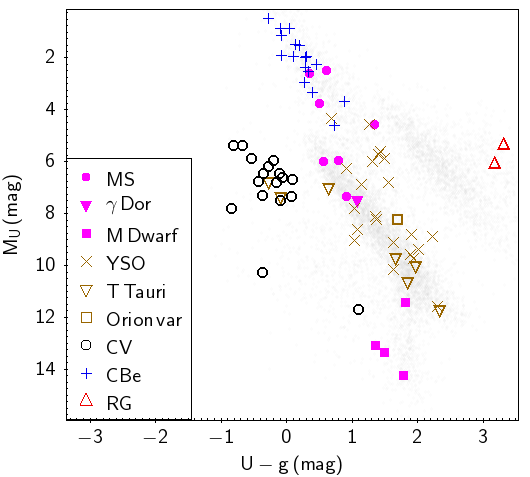}
	\includegraphics[width=\columnwidth, trim= 2mm 0mm 3mm 0mm]{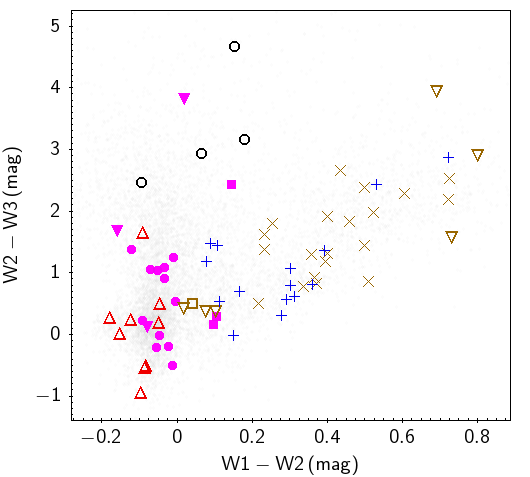}
	\caption{Location in \ucmd\ and \wccd\ of our sub-samples, with available photometric measurements in the corresponding bands. The colour (and shape)-code applied here is the same of Fig.~\ref{fig:classification_gaia}. The grey dots in the background depict the \ha-excess candidates found by FR21, included in UVEX or in WISE surveys.}
    \label{fig:sources_x_uvex_x_wise}
\end{figure}

Fig.~\ref{fig:sources_x_uvex_x_wise} shows the locations in the M\textsubscript{U} vs. g-r CMD\footnote{The zero-extinction assumption is also applied to obtain the absolute magnitude in the UVEX U band (M\textsubscript{U}).} (top panel) and in the W2-W3 vs. W1-W2 CCD (bottom panel) of all the objects with valid measurements in the corresponding bands.
Limiting the analysis to the \ha-emitting populations, a visual inspection of the three parameter spaces combined shows that Be stars and YSOs seem to be better separate in \ucmd\ than in the \gcmd. CVs are clearly bluer compared to the ZAMS in \ucmd\ than in the \gcmd\ while M-dwarfs appear to be equally well identifiable in \ucmd\ or the \gcmd. Future studies may be able to further leverage multi-band photometry to quantify the class separations of \ha-emitting sources. This may be particularly useful to optimise target selection in large spectroscopic surveys.

\onecolumn
\begin{longtable}[c]{| c | c | c | c | c | c | c | c |}
\caption{Results of our spectroscopic classification. For each target, the following entries are provided: \gaia\ DR2 SourceID (for almost all the sources in our dataset, the Source ID did not change between DR2 and EDR3. The only exception is \gaia\ DR2 4306883326285609728, that became \gaia\ EDR3 4306883326286990464); the \gaia\ DR2 barycentric Right Ascension and Declination, at epoch 2015.5; the distance, obtained inverting \gaia\ DR2 parallax; the apparent magnitude in IPHAS r band; the maximum significance (between the two provided by FR21) of being an \ha-excess source; the FWHM of the \ha\ line (``NA'', if said line is seen in absorption); our spectral classification.\label{tab: results}}\\

\hline
\shortstack{\textbf{SourceID} \\ (\gaia\ $DR2$) } & \shortstack{\textbf{RA}\\ (\textdegree)} & \shortstack{\textbf{Dec}\\ (\textdegree)} & \shortstack{\textbf{Distance}\\ (kpc)} & \shortstack{\textbf{r}\\(mag)} & \shortstack{\textbf{$\boldsymbol{n_{\sigma_{max[CMD,POS]}}}$}\\ \textbf{}} & \shortstack{\textbf{FWHM}\\ (\AA)} & \shortstack{\textbf{Class}\\ \textbf{}}  \\
\hline
\endfirsthead
\hline
\endhead

\hline
\endfoot
\endlastfoot

 183806256466977920 & 79.54732 & 35.13356 & 0.91(5) & 16.35 & 13.56 & 19(1) & CV \\
 \hline
 184238295817134208 & 82.51049 & 36.82756 & 0.62(2) & 14.69 & 6.42 & NA & MS \\
\hline
 185881786526045440 & 74.66526 & 35.09533 & 1.2(1) & 16.77 & 8.49 & 14.2(6) & CV \\
 \hline
 188400875040872832 & 76.02748 & 38.72527 & 2.8(3) & 13.66 & 5.76 & 9.14(7) & Be \\
 \hline
 202583509889708800 & 77.95731 & 44.28475 & 0.58(3) & 16.62 & 12.50 & NA & MS \\
 \hline
 207006089252554240 & 75.83720 & 47.08507 & 1.3(2) & 17.22 & 9.14 & 18.3(2) & CV \\
 \hline
 250899658585804544 & 59.41636 & 51.61552 & 0.506(7) & 13.60 & 9.38 & NA & MS \\
 \hline
 251693368542860288 & 59.30778 & 51.99074 & 1.19(6) & 15.61 & 11.32 & 10.15(8) & YSO \\
 \hline
 258089193320478208 & 67.22134 & 48.52566 & 1.9(1) & 15.44 & 7.74 & 11.17(4) & Be \\
 \hline
 259018688664493312 & 71.64038 & 48.96539 & 0.56(2) & 16.56 & 15.95 & 20.30(1) & CV \\
 \hline
 276781475113802496 & 62.89925 & 55.87940 & 4.7(7) & 14.82 & 7.7 & 10.15(7) & Be \\
 \hline
 422750061837839232 & 0.93751 & 58.46850 & 1.0(1) & 17.29 & 6.67 & NA & MS \\
 \hline
 424396924098269312 & 13.42935 & 57.96660 & 4.4(4) & 13.86 & 10.47 & NA & RG \\
 \hline
 425129194536945024 & 10.71283 & 58.32971 & 0.97(4) & 16.51 & 14.89 & NA & MS \\
 \hline
 425995304767655040 & 13.37703 & 59.19563 & 3.6(3) & 13.83 & 13.05 & NA & RG \\
 \hline
 427292453608805504 & 13.32775 & 60.77675 & 0.82(1) & 14.02 & 12.10 & NA & MS \\
 \hline
 427316058750043520 & 13.55209 & 60.94371 & 3.5(3) & 13.73 & 7.32 & NA & RG \\
 \hline
 427473838665897984 & 13.38334 & 61.84135 & 0.339(3) & 14.24 & 7.17 & NA & MS \\
 \hline
 428233081505186688 & 5.22420 & 59.16597 & 0.98(5) & 16.91 & 6.30 & NA & MS \\
 \hline
 463122170299415168 & 45.27165 & 60.53203 & 2.4(3) & 15.84 & 10.45 & 9.14(1) & YSO \\
 \hline
 464931073088436096 & 41.95254 & 60.96403 & 2.3(1) & 13.95 & 14.50 & 7.11(3) & Be \\
 \hline
 465712207381733888 & 41.97116 & 61.56827 & 6(1) & 14.37 & 9.23 & 7.11(3) & Be \\
 \hline
 510854753206972800 & 22.63292 & 62.35899 & 0.79(5) & 16.92 & 11.45 & 18.3(2) & CV \\
 \hline
 512016318518918400 & 26.73920 & 63.00894 & 6(1) & 15.64 & 16.33 & 13.20(4) & Be \\
 \hline
 514896283072146560 & 30.58104 & 63.37740 & 0.277(7) & 16.78 & 8.10 & 8.1(1) & M-Dwarf \\
 \hline
 522539950472429312 & 17.01763 & 61.66252 & 0.478(4) & 13.63 & 6.29 & NA & MS \\
 \hline
 522542802330612608 & 16.68316 & 61.54304 & 6(1) & 13.95 & 7.01 & NA & RG \\
 \hline
 522841835128697472 & 16.42042 & 62.10566 & 3.0(2) & 13.94 & 6.43 & NA & MS \\
 \hline
 524238661574474624 & 13.60656 & 64.62303 & 2.10(7) & 13.73 & 10.66 & NA & MS \\
 \hline
 524301986569104768 & 14.93122 & 64.91167 & 0.76(4) & 17.26 & 15.22 & 25.4(6) & CV \\
 \hline
 526118620296454528 & 13.55275 & 66.29011 & 0.657(8) & 13.92 & 7.58 & NA & MS \\
 \hline
 527026954341506688 & 8.95623 & 64.25796 & 3.7(6) & 16.44 & 15.88 & NA & RG \\
 \hline
 1823343535657240064 & 298.95041 & 20.02398 & 5.1(8) & 14.30 & 7.58 & NA & RG \\
 \hline
 1833363385113024896 & 300.48773 & 23.60346 & 1.8(2) & 17.04 & 7.69 & 17.1(4) & CV \\
 \hline
 1970080784065021184 & 317.62957 & 43.13736 & 1.8(2) & 16.69 & 7.79 & NA & MS \\
 \hline
 1971264992446842368 & 321.13415 & 45.06151 & 1.2(1) & 17.02 & 11.38 & 25.4(5) & CV \\
 \hline
 1971760631674787072 & 318.70753 & 44.74772 & 2.3(4) & 16.72 & 7.63 & 14.2(4) & CV \\
 \hline
 1978346259653262976 & 322.95547 & 48.06728 & 0.80(3) & 16.0 & 10.81 & 8.12(5) & YSO \\
 \hline
 1980077092801857664 & 331.73643 & 51.09846 & 1.09(9) & 16.68 & 22.70 & 23.3(2) & CV \\
 \hline
 1997162958039726080 & 348.79460 & 55.95792 & 2.5(3) & 15.79 & 8.97 & NA & MS \\
 \hline
 2007611926287679616 & 337.32027 & 57.08722 & 4.4(4) & 13.90 & 9.17 & 11.17(4) & Be \\
 \hline
 2015167735817519360 & 345.94170 & 61.81282 & 0.75(2) & 14.67 & 13.17 & 8.12(3) & YSO \\
 \hline
 2017612358184192768 & 359.83326 & 66.38672 & 0.79(2) & 14.63 & 15.15 & 7.11(1) & YSO \\
 \hline
 2018379134410850688 & 294.93998 & 22.08596 & 0.63(2) & 16.64 & 6.96 & 8.0(1) & T-Tauri \\
 \hline
 2021418317687630080 & 293.75924 & 24.77452 & 0.417(9) & 16.22 & 9.75 & 6.0(1) & M-Dwarf \\
 \hline
 2022704849371681024 & 290.84948 & 23.73654 & 0.53(1) & 15.56 & 6.63 & NA & Gamma Dor \\
 \hline
 2024391951234118912 & 292.20999 & 24.98576 & 2.1(3) & 16.75 & 8.75 & NA & Gamma Dor \\
 \hline
 2030374978109714688 & 298.23984 & 30.08825 & 1.42(5) & 15.41 & 12.01 & NA & MS \\
 \hline
 2033218590053883648 & 294.44810 & 31.67799 & 5(1) & 16.04 & 6.24 & NA & MS \\
 \hline
 2034706057485253376 & 297.86111 & 32.94033 & 1.03(7) & 17.20 & 6.80 & NA & MS \\
 \hline
 2034718873668327680 & 297.07516 & 32.64354 & 6(1) & 14.83 & 5.84 & NA & RG \\
 \hline
 2046886550363859968 & 295.31150 & 32.87483 & 3.8(5) & 16.08 & 8.31 & NA & MS \\
 \hline
 2054712221272378496 & 304.89408 & 33.57262 & 1.19(7) & 16.41 & 5.72 & NA & MS \\
 \hline
 2061420410434776192 & 305.63634 & 39.28241 & 0.79(1) & 13.81 & 11.67 & NA & MS \\
 \hline
 2066074436989021440 & 311.37173 & 41.05052 & 3.0(3) & 15.43 & 7.12 & 12.18(6) & YSO \\
 \hline
 2071104221657306496 & 309.20076 & 45.34096 & 0.93(3) & 15.94 & 13.20 & 5.08(3) & YSO \\
 \hline
 2162336164614235648 & 316.96769 & 44.09499 & 0.475(5) & 13.94 & 10.86 & 21.32(6) & CV \\
 \hline
 2162850736074155392 & 316.39512 & 46.37367 & 3.3(4) & 15.62 & 7.37 & NA & MS \\
 \hline
 2162947798009095808 & 312.86245 & 44.22094 & 0.79(4) & 16.93 & 20.98 & 6.09(5) & YSO \\
 \hline
 2164154516334658560 & 318.81208 & 46.45168 & 3.9(3) & 14.61 & 8.45 & NA & MS \\
 \hline
 2165559073718735872 & 316.57904 & 48.68376 & 3.3(5) & 15.92 & 13.61 & NA & RG \\
 \hline
 2166302068698392960 & 311.57080 & 45.42623 & 3.7(6) & 16.14 & 9.97 & 10.15(6) & YSO \\
 \hline
 2168759236665389312 & 315.91445 & 50.26468 & 0.59(1) & 15.69 & 10.23 & 7.11(6) & YSO \\
 \hline
 2170925927411884672 & 323.05921 & 49.74432 & 1.6(2) & 17.02 & 13.46 & 16.2(2) & CV \\
 \hline
 2171539832855206784 & 324.25968 & 51.53298 & 0.208(3) & 16.85 & 6.96 & 7.1(1) & M-Dwarf \\
 \hline
 2201164799270238976 & 334.85416 & 59.55555 & 3.4(4) & 15.07 & 16.14 & 8.12(1) & YSO \\
 \hline
 2207626251145725440 & 345.36781 & 62.93661 & 0.89(2) & 14.79 & 6.35 & 9.14(3) & YSO \\
 \hline
 2209703675287141504 & 356.09811 & 66.02327 & 1.40(8) & 15.89 & 8.65 & NA & MS \\
 \hline
 2210266522162289536 & 350.61836 & 66.02533 & 0.74(1) & 14.77 & 21.19 & 9.14(2) & YSO \\
 \hline
 3105714495538124288 & 101.47439 & -2.90572 & 6(1) & 14.87 & 12.27 & 12.05(3) & Be \\
 \hline
 3113703547027886720 & 104.11159 & 0.38287 & 1.4(2) & 17.31 & 11.52 & NA & MS \\
 \hline
 3119336516892417024 & 99.76717 & -0.80271 & 1.9(3) & 16.07 & 9.99 & 8.03(3) & YSO \\
 \hline
 3125759898182046080 & 102.12067 & 1.51568 & 3.7(4) & 13.84 & 7.67 & 10.04(2) & Be \\
 \hline
 3126136064298079616 & 102.23128 & 1.59818 & 4.8(8) & 13.9 & 13.62 & 8.03(2) & Be \\
 \hline
 3126247218051915264 & 103.83627 & 2.31710 & 0.72(6) & 16.52 & 6.70 & NA & Gamma Dor \\
 \hline
 3126354699614424064 & 102.17797 & 2.06890 & 1.3(2) & 16.77 & 6.91 & NA & Gamma Dor \\
 \hline
 3129126499706430976 & 103.45549 & 4.99618 & 4.5(7) & 14.06 & 4.97 & 9.03(2) & Be \\
 \hline
 3130722814493379968 & 100.68126 & 4.59085 & 0.275(9) & 16.75 & 9.55 & 6.0(1) & M-Dwarf \\
 \hline
 3132911705924538112 & 102.16806 & 6.64056 & 3.9(7) & 14.76 & 8.94 & 15.06(6) & Be \\
 \hline
 3324947321590659456 & 94.65241 & 6.94508 & 0.65(3) & 16.38 & 23.69 & 9.03(3) & T-Tauri \\
 \hline
 3326685615112100992 & 100.31993 & 9.45838 & 0.73(2) & 14.38 & 33.26 & 10.04(3) & T-Tauri \\
 \hline
 3326734470363384320 & 100.68635 & 9.98419 & 0.76(6) & 16.87 & 5.95 & 7.0(3) & T-Tauri \\
 \hline
 3326896854488117632 & 99.87232 & 9.72769 & 0.73(3) & 15.92 & 6.40 & 7.0(1) & T-Tauri \\
 \hline
 3326904585429362560 & 100.10781 & 9.84931 & 0.73(6) & 16.20 & 17.18 & 6.02(3) & T-Tauri \\
 \hline
 3326938262268528128 & 100.43208 & 10.13123 & 0.70(3) & 14.68 & 6.46 & 8.0(2) & Orion Var \\
 \hline
 3349903246243514752 & 88.17739 & 17.18745 & 0.77(4) & 15.64 & 6.07 & 8.03(4) & T-Tauri \\
 \hline
 3350768218295646336 & 100.70632 & 10.18121 & 0.469(9) & 14.01 & 5.80 & NA & Gamma Dor \\
 \hline
 3355257352475460480 & 98.03440 & 12.48440 & 6(1) & 13.99 & 8.46 & 8.03(1) & Be \\
 \hline
 3356132937390349824 & 99.12068 & 14.54367 & 1.2(2) & 16.09 & 7.86 & 27.1(3) & CV \\
 \hline
 3369781587542465280 & 96.72360 & 17.79049 & 5.1(8) & 14.37 & 7.41 & 9.04(3) & Be \\
 \hline
 3372309639651922432 & 97.35764 & 19.57940 & 3.0(3) & 15.33 & 7.34 & NA & MS \\
 \hline
 3376955492893794048 & 93.22946 & 22.17039 & 0.78(4) & 15.93 & 8.35 & 23.1(2) & CV \\
 \hline
 3424187076448918272 & 88.42817 & 21.87315 & 3.5(5) & 14.41 & 8.30 & 12.05(5) & Be \\
 \hline
 3444277112392781440 & 86.47305 & 29.61278 & 4.0(4) & 13.95 & 8.16 & 15.23(3) & Be \\
 \hline
 3448388529968087040 & 85.47474 & 32.60364 & 4.7(8) & 14.43 & 9.99 & 10.15(5) & Be \\
 \hline
 3451057044687431296 & 89.29311 & 32.65179 & 2.7(3) & 15.30 & 26.23 & NA & MS \\
 \hline
 3455444577118544768 & 85.92677 & 34.93837 & 4.2(6) & 14.75 & 6.41 & Inv\footnote{The spectrum of this object presents \ha-inversion, as shown in Fig.~\ref{fig:weird_spectra}.}. & Be \\
 \hline
 4259907860230511616 & 281.70494 & -1.61207 & 3.6(7) & 14.06 & 6.52 & 9.03(3) & YSO \\
 \hline
 4260073676017333248 & 281.49998 & -1.21673 & 3.0(5) & 15.12 & 8.02 & 11.04(3) & YSO \\
 \hline
 4261708203134890624 & 286.48896 & -2.63683 & 1.0(1) & 16.34 & 9.33 & 19.1(4) & CV \\
 \hline
 4261769672710353920 & 285.84803 & -2.29662 & 0.227(3) & 16.48 & 14.9 & 28.1(2) & CV \\
 \hline
 4265284712670427264 & 282.72070 & -1.15056 & 3.3(7) & 16.78 & 13.38 & NA & RG \\
 \hline
 4273688795363205760 & 277.83036 & 1.11545 & 0.452(7) & 14.61 & 7.25 & 8.03(3) & YSO \\
 \hline
 4278629893245072384 & 282.28770 & 1.42557 & 0.58(1) & 14.66 & 6.41 & 10.04(7) & YSO \\
 \hline
 4281419427268084480 & 286.68511 & 4.66668 & 0.71(3) & 15.40 & 9.3 & 10.04(7) & YSO \\
 \hline
 4282917305693991936 & 284.19650 & 7.01528 & 0.40(2) & 17.12 & 6.32 & 26.1(2) & CV \\
 \hline
 4283915662276175104 & 279.88333 & 4.99783 & 3.1(5) & 15.65 & 10.57 & NA & RG \\
 \hline
 4305857275780220032 & 286.25677 & 6.40762 & 2.3(4) & 16.94 & 5.98 & 9.04(6) & YSO \\
 \hline
 4306883326285609728 & 284.75531 & 7.12050 & 1.12(6) & 14.66 & 10.58 & NA & Gamma Dor \\
 \hline
 4309797032104899328 & 288.96905 & 11.20355 & 1.9(1) & 13.88 & 13.53 & 12.05(6) & YSO \\
 \hline
 4311951731334685440 & 283.65170 & 10.57241 & 0.98(3) & 14.32 & 7.16 & NA & MS \\
 \hline
 4311967365016126976 & 283.85071 & 10.91176 & 1.8(1) & 15.52 & 6.72 & NA & Gamma Dor \\
 \hline
 4316059300586640256 & 291.39037 & 12.68881 & 3.4(6) & 15.05 & 6.23 & 16.06(9) & YSO \\
 \hline
 4514006371833731456 & 286.21437 & 17.35407 & 1.8(2) & 16.62 & 9.68 & NA & Gamma Dor \\
 \hline

\end{longtable}
\twocolumn


\bsp	
\label{lastpage}
\end{document}